\newcommand*{\FinalTypeSetting}{}

\ifdefined\FinalTypeSetting
\documentclass[prc,reprint,showpacs,superscriptaddress,nofootinbib]{revtex4-1}
\else
\documentclass[prc,preprint,showpacs,notitlepage,floatfix,superscriptaddress]{revtex4-1}
\fi
\usepackage{latexsym}
\usepackage[utf8]{inputenc}
\usepackage{graphicx}
\usepackage{color}
\usepackage{units}
\usepackage{xcolor,colortbl}
\usepackage{amsmath,amssymb,bm}
\usepackage{amsfonts}
\usepackage{comment}
\usepackage[caption=false]{subfig}
\usepackage{hyperref}
\hypersetup{
    colorlinks,
    linkcolor={red!50!black},
    citecolor={blue!50!black},
    urlcolor={blue!80!black}
}
\usepackage{longtable}
\ifdefined\FinalTypeSetting
\usepackage[disable]{todonotes}

\else
\usepackage{todonotes}
\fi

  \newcommand{\co}{(Color online). } 
                                
\graphicspath{{fig-article/}}

\begin{document}
\listoftodos
\title{Uncertainty quantification for proton-proton fusion in chiral
  effective field theory}

\author{B.~Acharya} 
\affiliation{Department of Physics and Astronomy,
University of Tennessee, Knoxville, TN 37996, USA}

\author{B. D. Carlsson}
\affiliation{Department of Physics, Chalmers University of Technology, SE-412 96 G\"oteborg, Sweden}

\author{A. Ekstr\"om}
\affiliation{Department of Physics and Astronomy,
University of Tennessee, Knoxville, TN 37996, USA}
\affiliation{Physics Division,
Oak Ridge National Laboratory, Oak Ridge, TN 37831, USA}
\affiliation{Department of Physics, Chalmers University of Technology, SE-412 96 G\"oteborg, Sweden}

\author{C. Forss\'en}
\affiliation{Department of Physics, Chalmers University of Technology, SE-412 96 G\"oteborg, Sweden}

\author{L.~Platter}
\affiliation{Department of Physics and Astronomy,
University of Tennessee, Knoxville, TN 37996, USA}
\affiliation{Physics Division, Oak Ridge National Laboratory, Oak Ridge, TN 37831, USA}

\date{\today}

\begin{abstract}

  We compute the $S$-factor of the proton-proton ($pp$) fusion
  reaction using chiral effective field theory ($\chi$EFT) up to
  next-to-next-to-leading order (NNLO) and perform a rigorous
  uncertainty analysis of the results. We quantify the uncertainties
  due to (i) the computational method used to compute the $pp$ cross
  section in momentum space, (ii) the statistical uncertainties in the
  low-energy coupling constants of $\chi$EFT, (iii) the
  systematic uncertainty due to the $\chi$EFT cutoff, and (iv) 
  systematic variations in the database used to calibrate the
  nucleon-nucleon interaction. We also examine the robustness of the
  polynomial extrapolation procedure, which is commonly used to extract the threshold
  $S$-factor and its energy-derivatives. By performing a statistical
  analysis of the polynomial fit of the energy-dependent $S$-factor at
  several different energy intervals, we eliminate a systematic
  uncertainty that can arise from the choice of the fit interval in
  our calculations. In addition, we explore the statistical correlations between the
  $S$-factor and few-nucleon observables such as the binding energies
  and point-proton radii of $^{2,3}$H and $^{3}$He as well as the
  $D$-state probability and quadrupole moment of $^2$H, and the
  $\beta$-decay of $^{3}$H. We find that, with the state-of-the-art optimization of the nuclear Hamiltonian, 
  the statistical uncertainty in the threshold $S$-factor cannot be reduced beyond 0.7\%.

\end{abstract}

\maketitle

\section{Introduction
\label{sec:introduction}} 
In main sequence stars such as the Sun, the conversion of hydrogen to
helium proceeds predominantly through the $pp$ chain, which is
primarily triggered by the weak $pp$-fusion process
~\cite{Adelberger:1998qm,Adelberger:2010qa},
\begin{equation}
\label{eq:reaction}
 p + p \rightarrow d + e^+ + \nu_e.
\end{equation} 
The accurate determination of the rate of this reaction is a critical
ingredient for our understanding of many stellar processes. The
reaction rate is conventionally parametrized in terms of the
$S$-factor, which is related to the cross section by
\begin{equation}
S(E) = \sigma(E) E e^{2\pi\eta}, 
\end{equation}
where $E$ is the center of mass energy and
$\eta = \sqrt{m_p/E}\,\alpha/2$ is the Sommerfeld parameter. This
cross section can only be measured in experiments performed at rather
high energies in order to overcome the Coulomb barrier between the
protons. Extrapolations of $S(E)$ to relevant energy domains,
$E < 10$~keV, which is where the Gamow peak of the Sun lies, yield
extremely large uncertainties. Therefore, we are forced to rely on
theoretical calculations to provide a precise prediction. This
situation turns the accompanying uncertainty analysis into an absolute
necessity. However, the
quantification of theoretical uncertainties, which can have many
different origins, is a very difficult task and a rigorous uncertainty analysis is still lacking.
It is the purpose of this Letter to significantly advance the
state-of-the-art of theoretical uncertainty quantification for the
$pp$-fusion process.

Although calculations of $S(E)$ can be performed at first order in the
weak coupling constant, they still require a detailed knowledge of the
nuclear interaction, which has to be treated
nonperturbatively. Potential models therefore provided the first
insights into this
process~\cite{Bethe:1938yy,Bahcall:1968wz,Kamionkowski:1993fr,Schiavilla:1998je}. Still, 
obtaining reliable uncertainty estimates has always proven to be hard
with such models for the nucleon-nucleon interaction. 
Effective field theories (EFTs) are systematic low-energy expansions
in a small parameter and are capable to provide an estimate of the 
inherent systematic error. An effort to provide uncertainty estimates was first carried out in~\cite{Park:1998wq,Park:2002yp} using the so-called
hybrid approach in which current operators were obtained from 
$\chi$EFT and wave functions from phenomenological potentials. The
first complete $\chi$EFT calculation of the $S$-factor was carried out
by Marcucci {\it et al.}~\cite{Marcucci:2013tda}, who obtained
\begin{equation}
  \label{eq:SfactorMarcucci}
  S=(4.030 \pm 0.006)\times 10^{-23}~{\rm MeV~fm}^2~,
\end{equation}
with an uncertainty (see below) seven times smaller than that of the
previously recommended value~\cite{Adelberger:2010qa}
\begin{equation}
S=(4.01 \pm 0.04)\times 10^{-23}~{\rm MeV~fm}^2~.
\end{equation}
Pionless EFT
calculations~\cite{Kong:2000px,Butler:2001jj,Ando:2008va,Chen:2012hm},
which use only the nucleons as explicit degrees of freedom, have
obtained consistent values, albeit with slightly larger theory errors.

Since uncertainty analysis was not the main objective of Ref.~\cite{Marcucci:2013tda}, 
their error estimates were based on simple assumptions. The current operators and the wavefunctions 
were calculated to high precision --- $\mathcal{O}(Q^3)$ and $\mathcal{O}(Q^4)$, respectively --- 
by employing $\chi$EFT interactions widely used in the literature. 
The uncertainty was then estimated from the range of $S$-factor values obtained by using two
different short-distance regulators in the two-body current and the potential. 
Thus, the error reported in Ref.~\cite{Marcucci:2013tda} only reflects the 
resulting spread in the $S$-factor at a mixed order in the EFT expansion.

In this Letter we study the $pp$-fusion process in $\chi$EFT at
next-to-next-to-leading order (NNLO). In particular, we perform an
accompanying uncertainty quantification that builds on recent progress
in mathematical optimization and statistical analysis of chiral
nuclear forces and \textit{ab initio} nuclear
theory~\cite{Ekstrom:2013kea,navarro2014,Ekstrom2015,navarro2015b,Carlsson:2015vda}. We
quantify both statistical uncertainties, associated with the
determination of the relevant low-energy constants (LECs), as well as
several systematic ones, related to the computational method and to the
chiral expansion. In this process we aim for consistency, e.g. by
regulating the weak-current operator of the $pp$-fusion process in a
manner that is consistent with the $\beta$-decay used in the fit of
the nuclear potential.

This paper is organized as follows. In Sec.~\ref{sec:formalism}, we
discuss the weak current operator and the initial and final state wave
functions. In Sec.~\ref{sec:calculations}, we present an analysis of
the uncertainties in the $S$-factor calculation.  In
Sec.~\ref{sec:results-discussion}, we present our final results along
with a discussion.

\section{Formalism}
\label{sec:formalism}
The cross section for the reaction in Eq.~\eqref{eq:reaction} can be
written in the center of mass frame as
\begin{align}
 \label{eq:crossec}
 \sigma(E) = \int
  & \frac{\mathrm{d}^3p_e}{(2\pi)^3}\frac{\mathrm{d}^3p_\nu}{(2\pi)^3}\frac{1}{2E_e} \frac{1}{2E_\nu}\nonumber\\
  & 2\pi\delta\left(E+2m_p-m_d-\frac{q^2}{2m_d}-E_e-E_\nu\right)\nonumber\\
  & \frac{1}{v_{rel}}\,F(Z,E_e)\,\frac{1}{4}\sum \vert \langle f \vert \hat H_W \vert i \rangle  \vert^2,
\end{align}
where $p_{e,\nu}$ are the positron and neutrino momenta, $E_{e,\nu}$
their energies, $m_d$ is the deuteron mass, $v_{rel}$ is the $pp$
relative velocity, and $q$ is the momentum of the recoiling
deuteron. The function $F(Z,E_e)$ accounts for the distortion of the
positron wave function due to the Coulomb field of the deuteron. Its
classical expression, which can be found in
Ref.~\cite{Feenberg:1950ft}, is increased by 1.62\% due to radiative
corrections~\cite{Kurylov:2002vj}. The summation runs over the spin
projections of all the initial and the final state particles. The
initial state $\vert i \rangle$ and the final state $\vert f \rangle$
are direct products of leptonic and nuclear states.  At nuclear
energies, the weak interaction Hamiltonian can be written in terms of
the leptonic and the nuclear weak current operators as
\begin{equation}
\label{eq:weak_current_operator}
 \hat H_W = \frac{G_V}{\sqrt{2}}\int\mathrm{d}^3x \left[j_\mu(\mathbf{x}) {J^\mu}^\dagger(\mathbf{x})+\mbox{h.c.}\right],
\end{equation}
where $G_V$ is the vector coupling constant. 
The matrix element of the leptonic weak current operator $j^\mu$ between the leptonic
wave functions is $l^\mu\,e^{-i \mathbf{q}\cdot\mathbf{x}}$, where
$l^\mu$ satisfies
\begin{align}
\label{eq:leptonspinsum}
 \sum l^\sigma {l^\tau}^\ast = 8 \big( & \,{p_\nu}^\sigma\,{p_e}^\tau+{p_\nu}^\tau\,{p_e}^\sigma\nonumber\\
 & -g^{\sigma\tau}{p_\nu}^\rho\,{p_e}_\rho - i \epsilon^{\rho\sigma\delta\tau}{p_\nu}_\rho\,{p_e}_\delta\,\big),
\end{align}
with the summation running over the spin projections of the
leptons. Note that Eqs.~\eqref{eq:crossec} and
\eqref{eq:leptonspinsum} look different from the ones in
Ref.~\cite{Walecka:1995mi,Marcucci:2013tda} because we include the $2E_e$
and $2E_\nu$ denominators in Eq.~\eqref{eq:crossec} to make the phase
space volume element manifestly Lorentz invariant.

The nuclear wave functions are calculated up to NNLO in $\chi$EFT. The
nuclear weak current operators are consistently derived from the same
effective Lagrangian up to the same order in $Q$,
i.e. $\mathcal{O}(Q^3)$. In this approach, the strong interaction dynamics between
the nucleons involved is based on the same theoretical grounds as
their coupling to the leptonic current.

\subsection{Nuclear weak current operators}
\label{sec:nuclear-weak-current}
The current operators for charge-changing weak interactions were
derived in $\chi$EFT in
Refs.~\cite{Park:1995pn,Park:1998wq,Park:2002yp,Song:2008zf}.

The one-body (1B) operators that give non-vanishing matrix elements
between an $S$-wave $pp$ wave function and the $S$- and $D$-wave
configurations of the deuteron include, up to $\mathcal{O}(Q^3)$, the
Gamow-Teller operator,
\begin{equation}
 \mathbf{J}^-_\mathrm{GT} = - g_A \displaystyle\sum_l \tau_l^-\bm{\sigma}_l,
\end{equation} at leading order and the ``weak-magnetism'' operator
\begin{equation}
 \mathbf{J}^-_\mathrm{WM} = \displaystyle\sum_l \tau_l^- i \frac{\mu_V}{2m_N} \mathbf{q}\times\bm{\sigma}_l,
\end{equation}
which is $\mathcal{O}(Q)$. Here the index $l$ runs over both
nucleons, $g_A$ is the axial vector coupling constant, $\tau_l^-$ is
the isospin lowering operator $(\tau_l^x-i\tau_l^y)/2$. Formally,
there are additional operators~\cite{Park:2002yp,Menendez:2011qq} that
enter according to the $\chi$EFT power counting scheme~\footnote{Note
  that Refs.~\cite{Park:2002yp} and \cite{Menendez:2011qq} use
  slightly different power-counting
  schemes. Ref.~\cite{Menendez:2011qq} uses
  $q/m_N\sim\mathcal{O}(Q^2)$, which is more appropriate for the
  energy regime they consider.}. The matrix elements of those
operators are, however, kinematically suppressed for the extremely
small proton energies being considered here. The $\chi$EFT 1B
currents operators used here have the same structure as the ones
obtained phenomenologically in earlier studies (see, $e.g.$
Ref.~\cite{Tomoda:1990rs}).

The expression for the axial-vector two-body (2B) current, which is
$\mathcal{O}(Q^3)$, reads
\begin{align}
\mathbf{J}^-_\mathrm{2B} = -\frac{g_A}{2 F_\pi^2}\bigg\{&\frac{1}{m_\pi^2+\mathbf{k}^2}\bigg[ -\frac{i}{2m_N} \tau_\times^-\,\mathbf{p} (\bm{\sigma}_1-\bm{\sigma}_2) \cdot\mathbf{k} \nonumber\\
& + 4 c_3 \, \mathbf{k}\,[\mathbf{k}\cdot\displaystyle\sum_l \tau_l^-\bm{\sigma}_l] \nonumber\\
& + \left( c_4 +\frac{1}{4m_N}\right) \tau_\times^-\,\mathbf{k}\times[\bm{\sigma}_\times\times\mathbf{k}] \bigg] \nonumber\\
& + 4 d_1 \displaystyle\sum_l \tau_l^-\bm{\sigma}_l + 2 d_2\,\tau_\times^-\bm{\sigma}_\times\bigg\}.
\end{align}

Here
$\mathbf{k}=(\mathbf{p}^\prime_2-\mathbf{p}_2-\mathbf{p}^\prime_1+\mathbf{p}_1)/2$
and
$\mathbf{p}=(\mathbf{p}_1+\mathbf{p}^\prime_1-\mathbf{p}_2-\mathbf{p}^\prime_2)/4$,
where $\mathbf{p}_l$ ($\mathbf{p}^\prime_l$) is the nucleon momentum
in the initial (final) state, $\tau_\times^- = (\tau_1\times\tau_2)^x
- i (\tau_1\times\tau_2)^y$, $\bm{\sigma}_\times
=\bm{\sigma}_1\times\bm{\sigma}_2$ and $F_\pi$ is the pion decay
constant. The constants $d_1$ and $d_2$ that appear in the short-range
2B current are constrained by the Pauli principle such that only one
linear combination, $d_1 + 2 d_2 = c_D/(g_A\Lambda_\mathrm{EFT})$
enters. The constants $c_3$ and $c_4$, which accompany the one-pion
exchange current, also appear in the $NN$ and $\pi N$ interactions, as
well as along with $c_D$ in the $NNN$ interaction. For the current, it
is customary to define the counterterm
\begin{equation}
 \hat d_R = \frac{1}{m_N} \frac{c_D}{g_A\,\Lambda_\mathrm{EFT}} + \frac{1}{3m_N} (c_3 + 2 c_4)+\frac{1}{6}
\end{equation}
to parametrize the strength of the meson-exchange current.

\subsection{Nuclear wave functions}
\label{sec:wavefunctions}

The NNLO momentum-space potentials that we employ are non-local. We
obtain the $S$-state ($L=0$) and $D$-state ($L=2$) components of the
deuteron wave function in coordinate space by diagonalizing in a
harmonic-oscillator basis. We find that it is necessary to correctly
    reproduce the deuteron wave function beyond 50~fm in order
    to achieve infrared convergence of the radial integrals.

This bound-state problem is easily solved equally well in either
momentum space or coordinate space. Unfortunately, this dual approach
is not as trivial for the relative $pp$-scattering wave function,
$\psi(r;E)$ as the presence of the long-ranged Coulomb potential
implies that the momentum space representation of the scattering
potential becomes singular for equal incoming and outgoing relative
momenta. In order to facilitate a numerical solution we follow the
prescription of Vincent and Phatak~\cite{Vincent:1974} and introduce a
cutoff radius $R_{c}$. For radii $r<R_{c}$, the Coulomb potential is
well-defined in momentum space. In this region we can therefore apply
ordinary methods to find $\psi(r<R_{c};E)$. Furthermore, if we choose
$R_{c} > 35$~fm, such that the short-ranged nuclear interaction
becomes negligible, we can smoothly match the solution at $r=R_{c}$ to
the asymptotic Coulomb wave function,
\begin{align}
\displaystyle\lim_{r\rightarrow\infty} \chi_0(r;E) = & \cos\delta_0\,F_0(r\sqrt{m_pE}) \nonumber\\
& +\sin\delta_0\,G_0(r\sqrt{m_pE}),
\end{align}
where $\delta_0$ is the $S$-wave phase shift with respect to the
nuclear and Coulomb potential, $F_0$ and $G_0$ are, respectively, the
regular and the irregular Coulomb wave functions, and $\chi_0(r;E)$ is
the radial $S$-wave $pp$ wave function, which, upon ignoring higher
partial wave contributions, is related to $\psi (r;E)$ by
\begin{equation}
 \psi (r;E) = \sqrt{\frac{2}{m_pE}}\,e^{i\delta_0}\frac{1}{r}\,\chi_0(r;E).
\end{equation}

The $pp$ wave function obtained is correct to NNLO in $\chi$EFT and
first order in the electromagnetic coupling constant, $\alpha$. We do
not include higher-order electromagnetic contributions explicitly in
this work. These electromagnetic interaction terms mainly lower the
central value of the $pp$-fusion cross
section~\cite{Schiavilla:1998je,Park:1998wq,Marcucci:2013tda}, as
discussed in Sec.~\ref{sec:results-discussion}, and leave the
corresponding error due to uncertainties in the description of the
strong nuclear force unchanged.

\subsection{Radial matrix elements}

The earliest calculations of the $S$-factor were performed using the
Gamow-Teller operator only. In this approximation, the $S$-factor is
proportional to $\Lambda^2 (E)$, the square of the overlap between the
$pp$ wave function at energy $E$ and the deuteron wave
function, given by 
\begin{equation}
\Lambda (E) = {\left(\frac{\gamma^3}{2m_pE}\right)}^{\frac{1}{2}} \, 
\frac{e^{i\delta_0}}{C_0} \int_0^\infty \mathrm{d}r \, u_d(r) \, \chi_0(r;E),
\end{equation}
where $C_0^2 = 2\pi\eta/(e^{2\pi\eta}-1)$. However, the $S$-factor we
calculate includes deuteron recoil effects and meson-exchange current
contributions. The former modifies not only the phase space but also
the matrix element in Eq.~\eqref{eq:crossec}. The meson-exchange
current, which turns out to be the dominant correction, is
conventionally quantified as $\delta_{\rm 2B}$, defined as the ratio
of the matrix element of $\mathbf{J}^-_\mathrm{\rm 2B}$ to that of
$\mathbf{J}^-_\mathrm{GT}$~\cite{Park:2002yp}. While the dependence of
$\Lambda(E)$ on the LECs of $\chi$EFT is solely through the $NN$
interaction in the initial and the final state wave functions,
$\delta_{\rm 2B}$ depends explicitly on many of the LECs through the
current operator, and therefore deserves a careful scrutiny in spite
of its relatively small contribution to the overall value of the
$S$-factor.

\section{\label{sec:calculations}Calculations}
In this section we compute the quantities that are relevant for
describing low-energy $pp$ fusion. In addition, we will
explore several sources of statistical and systematic uncertainties.

\subsection{Axial-vector coupling constant}
The $S$-factor depends predominantly on $g_A$ explicitly, by the
relation $ S\propto {g_A}^2$. The most recent world average value for
$g_A$, calculated by the Particle Data Group (PDG), is
$1.2701(25)$~\cite{1674-1137-38-9-090001}.  The values obtained in some
of the recent
experiments~\cite{PhysRevLett.105.181803,PhysRevLett.88.211801} are
higher than the world average.  For our uncertainty analysis, we use
the PDG recommended value which in turn is very close to the $1.2695(29)$ value used by Marcucci {\it et al.} in
Ref.~\cite{Marcucci:2013tda}.

\subsection{Chiral interactions and currents}
In this work we employ a large set of 42 different NNLO$_{\rm sim}$
interactions~\cite{Carlsson:2015vda} to describe the nuclear force. 
This family of potentials are derived from $\chi$EFT up to NNLO. 
However, each of them was constructed using one of the seven different 
regulator cutoffs $\Lambda=[450,475, 500, 525, 550, 575, 600]$ MeV, 
and their LECs were constrained using one of six different truncations of the 
maximum scattering energy in the
world database of $NN$ scattering cross sections. 
It is also particularly useful that
the covariance matrix of the LECs have been precisely determined
for each NNLOsim interaction. Therefore, utilizing this large set
of systematically optimized interactions, with known statistical
properties, allows us to better uncover systematic uncertainties
and explore the statistical uncertainties and correlations in the
pp-fusion process. It should be pointed out that an equally good
description of the fit data is attained with all NNLO$_{\rm sim}$
interactions, and the numerical value of all constrained LECs are
of natural order. Thus, all NNLO$_{\rm sim}$ interactions are
expected to perform very well in the few-nucleon sector of
nuclear physics. In detail, the strengths of all relevant LECs
were \textit{simultaneously} optimized to reproduce the selected
$NN-$ scattering cross sections, $\pi N-$scattering cross
sections, the binding energies and charge radii of $^{2,3}$H and
$^{3}$He, the quadrupole moment of $^{2}$H, as well as the 
comparative $\beta$-decay half-life of $^{3}$H. The systematic
uncertainty stemming from the excluded higher-order contributions
in the chiral expansion was also accounted for in the
determination of the LECs. A detailed expose of the optimization
protocol is given in Ref.~\cite{Carlsson:2015vda}.

Furthermore, and similarly to the work in Ref.~\cite{Park:2002yp}, we
regulate the ultraviolet behavior of the matrix elements of all
current operators using a Gaussian regulator on the form
$\exp[-\mathbf{k}^2/2\Lambda_\mathrm{EFT}^2]$, where $\mathbf{k}$ is
defined in Sec.~\ref{sec:nuclear-weak-current}. The $NN$-sector of the
NNLO$_{\rm sim}$ potentials is regulated in similar way;
$\exp[-(p/\Lambda_{\rm EFT})^{2n}]$, where $n=3$ and $p$ is the
relative momentum of the two interacting nucleons. We use the same
value for the cutoff $\Lambda_{\rm EFT}$ in the currents as we use in
the interactions.
\subsection{Central values}
The family of 42 optimized $\chi$EFT interactions at NNLO is employed
to construct an averaged central value for each one of the computed
quantities using the arithmetic mean of the separate calculations. The
magnitudes of the statistical uncertainties in each calculation are
nearly identical. Thus it is not necessary to explore weighted average
schemes.

The energy dependence of the astrophysical $S$-factor, is usually
parametrized using a polynomial expansion
\begin{equation}
S(E) = S(0) + S'(0)E + S''(0) E^2/2 + \ldots
\end{equation}
A second- or third-order polynomial is most common. Higher-order
polynomials turn out to be ill-conditioned extrapolants attributed
with large statistical uncertainties. The normalized $n$-th order
derivative $S^{n}(0)/S(0)$ often serve as input to e.g. neutrino flux
computations using solar models. We fit theoretical $S(E)$-values to a
third-order polynomial across an energy range $E=1-30$ keV. From this
we extract the $S$-factor at zero energy.

From the results obtained with the different interactions, we extract
the following averaged central values: $S(0)=4.081 \times 10^{-23}$
MeV fm$^2$, $\Lambda^2(0)=7.087$, $\delta_{\rm 2B}=0.43$\%,
$S'(0)/S(0)=10.84~\mathrm{MeV}^{-1}$, and $S''(0)/S(0)=317.8~\mathrm{MeV}^{-2}$. 

\subsection{Uncertainty analysis}
\label{sec:uncertainty-analysis-1}
We start by considering systematic uncertainties associated with the
methods used to compute the cross section and to extrapolate to zero
energy.  

A polynomial fit of $S(E)$ is uninformed by the underlying
physics of the $pp$ fusion process. The polynomial fit and
subsequent extrapolation to zero energy will depend on the limits
of the fit-interval $I_E = [E_{\rm min}, E_{\rm max}]$, for which
it is not evident a priori what limits to use. The numerical
precision of our computational code allows us to safely set
$E_{\rm min} = 1$~keV. To find the optimal $E_{\rm max}$, we
construct a penalty function $\chi^{2}$ that describes how well a
polynomial function $f^{\rm fit}$ describes the correponding $\chi$EFT
predictions $f^{\rm calc}$ in $I_{E}$,

\begin{equation}
\chi^2 = \sum_{i \in I_{E}} (1 - f_i^{\rm
  fit}/f_{i}^{\rm calc})^2~,
\end{equation}

and minimize this with respect to $E_{\rm max}$. Note that we
have also used the shorthand $f_{i}$ to denote $f(E_i)$, and
evaluate energies in $I_{E}$ at 1~keV intervals. Overall, we find
that a cubic polynomial fits the $\chi$EFT predictions better than
a quadratic polynomial.

\begin{figure}
    \centering

    \includegraphics[width=\columnwidth]{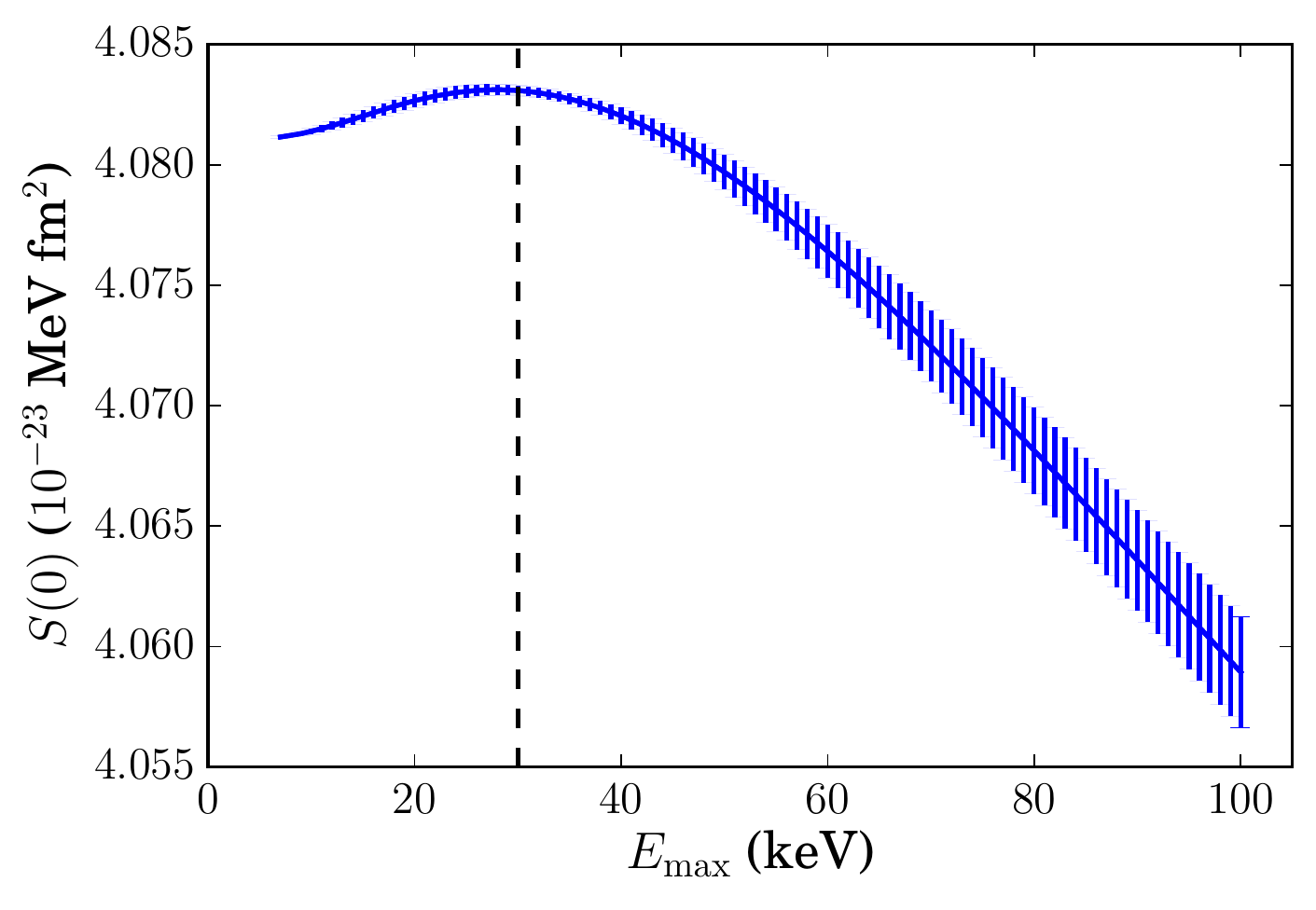}
    \caption{\label{fig:chi2_vs_Emax}
      Extrapolated $S(0)$-values using the cubic polynomial fit. The
      vertical error bars indicate the statistical uncertainty due to
      uncertainties in the polynomial fit coefficients. Clearly,
      they increase rapidly for $E_{\rm max} \gtrsim 30$ keV. }
\end{figure}

For a cubic polynomial, and by varying $E_{\rm max}$ between 4-100
keV, we find that $E_{\rm max} = 30$ keV is optimal. For this choice,
the statistical uncertainties of $S(0), S'(0)/S(0), S''(0)/S(0)$, due
to the polynomial fit, attain a minimum. The extrapolated
$S(0)$-values, as a function of $E_{\rm max}$ are shown in
Fig.~\ref{fig:chi2_vs_Emax}.
The central value for $S(0)$ decreases and the corresponding
statistical uncertainty of the polynomial fit increases when $E_{\rm
  max} \gtrsim 30$ keV. The same trend is observed for all 42 NNLO
interactions. The statistical error of the fit is ten times larger at
$E_{\rm max} = 100$ keV than it is at $E_{\rm max} = 30 $
keV. Therefore, we choose to fit all computed $S(E)$-values to a third
order polynomial across the energy interval $I_E = [1,30]$ keV. The
resulting statistical uncertainty for $S(0)$ that comes from the 
cubic polynomial fit is $0.0002 \times 10^{-23}$ MeV fm$^{2}$, which
corresponds to a relative uncertainty of $\sim 0.005\%$. Obviously,
the systematic uncertainty due to a possibly ill-determined $E_{\rm
  max}$ could be much larger, typically $\sim 1\%$. In order to
reliably extract the derivatives of the S-factor with respect to the
energy, a sufficiently large fit interval has to be chosen. However,
if the fit interval is made larger than approximately 30~keV the
necessity of including $P$-waves in the incoming channel becomes
apparent in Fig.~\ref{fig:chi2_vs_Emax}. The statistical
uncertainties, due to the polynomial fit, for the first and second
logarithmic derivatives of $S(0)$ are 0.02\% and 1.18\%,
respectively. The polynomial fit uncertainties for remaining
quantities are negligible.

In addition, we varied the Vincent-Phatak (VP) matching radius $R_{\rm cut}$
between 5-50 fm to extract uncertainties related to our numerical
approach to $pp$ scattering.

We find that $R_{\rm cut} = 35$~fm provides robust solutions, and we
conclude that the uncertainty in $S(0)$ due to the Vincent-Phatak
procedure is $ 0.002 \times 10^{-23}$ MeV fm$^2$, which corresponds to
a relative uncertainty of $\sim 0.05\%$.

The three remaining sources of uncertainties that we explore are
related to the $\chi$EFT description of the $pp$-fusion
process. First, there exists a statistical uncertainty due to the
non-zero variances of the LECs in the NNLO$_{\rm sim}$
interactions. Second, the actual choice of input $NN$ data is not
uniquely defined. Indeed, it is not clear what maximum kinetic energy
$T_{\rm Lab}$ one should consider in the selection of scattering
data. This ambiguity gives rise to a systematic
uncertainty. Third, variations in the regulator cutoffs, $\Lambda_{\rm
  EFT}$, of the $\chi$EFT description of the interaction and the
currents will induce another systematic uncertainty.

In the process of propagating statistical uncertainties we employ
well-founded methods for error propagation (see
Ref.~\cite{Carlsson:2015vda} for details). We also calculate the
corresponding statistical covariance matrix of the observables and
quantities that we compute. To find the covariance $\rm{Cov}(A,B)$
between two observables, $A$ and $B$, due to the statistical
covariances $\rm{Cov}(\vec \alpha)$ of the LECs $\mathbf{\vec \alpha}$, 
we employ the linear approximation
\begin{equation}
\rm{Cov}(A,B) = \mathbf{J}^{T}_{A}\rm{Cov}(\tilde{\alpha})\mathbf{J}_{B}
\end{equation}
where $J_{A}$ is the Jacobian vector of partial derivatives
$J_{A,i}=\frac{\partial A}{\partial \tilde{\alpha}_i}$, and similarly for
$J_{B}$. We compute $pp$-fusion in the $S$-wave channel only, thus the
relevant LECs are
$\tilde{\alpha} =
(c_1,c_3,c_4,\tilde{C}_{^1S_0}^{pp},\tilde{C}_{^3S_1},C_{^1S_0},C_{^3S_1},C_{^3S_1-^3D_1},c_D)$.
Further, we define the uncertainty in any observable $A$, due to the
uncertainties in the LECs, as
\begin{equation}
\sigma_{A} = \sqrt{\rm{Cov}(A,A)}.
\end{equation}
We use the covariance matrices $\rm{Cov}(\vec \alpha)$ from
Ref.~\cite{Carlsson:2015vda}, and obtain the necessary derivatives
$\frac{\partial A}{\partial \tilde{\alpha}_i}$ via a straightforward univariate
spline fit to 10 function evaluations of the observable $A$; changing
the LEC $\tilde{\alpha}_i$ in the neighborhood of its optimal value. We
benchmarked this spline approximation using the known derivative
values of the deuteron binding energy~\cite{Carlsson:2015vda}, and
found that it was accurate to at least $\sim 0.001 \%$.

From this procedure we find that the error in $S(0)$ due to the
statistical uncertainties in the LECs is $0.009 \times 10^{-23}$ MeV
fm$^{2}$, which implies a relative uncertainty of $\sim 0.2\%$. This
result is very stable for each of the simultaneously optimized NNLO
interactions. Propagating also the statistical errors of $g_A$
increases this statistical uncertainty to $0.019 \times 10^{-23}$ MeV
fm$^2$. The statistical uncertainty of $G_V$ has a negligible impact
on the statistical uncertainty of $S(0)$. Also, the statistical
uncertainties of all LECs have negligible impact on the logarithmic
derivatives of the zero-energy $S$-factor. The relative uncertainties
are; $0.05\%$ in $S'(0)/S(0)$ and slightly more , $0.2\%$, in
$S''(0)/S(0)$. Furthermore, the derivatives of $S(0)$ are very
insensitive to changes in the cutoffs $\Lambda_{\rm EFT}$ and
$T_{\rm Lab}^{\rm max}$. The relative uncertainty of the squared
overlap of the deuteron and proton-proton wavefunctions, $\Lambda^2$,
is $\sim 0.05\%$, and this error is dominated by the LEC
variations. The $\Lambda^2$ is only somewhat sensitive to variations in
$\Lambda_{\rm EFT}$ and $T_{\rm Lab}^{\rm max}$. For all 42 NNLO
interaction that we employed, we only observed
$\Lambda^2$ in the range $7.064-7.101$. Not surprisingly, the corresponding
uncertainty and variation in $\delta_{\rm 2B}$, which is more
sensitive to short distance physics, is much larger. The typical
relative uncertainty due to variations in the LECs is $\sim 5\%$.
Furthermore, $\delta_{\rm 2B}$ varies between 0.30\% and 0.52\% for
all NNLO interactions we explored.

\begin{figure}
    \centering
    \includegraphics[width=\columnwidth]{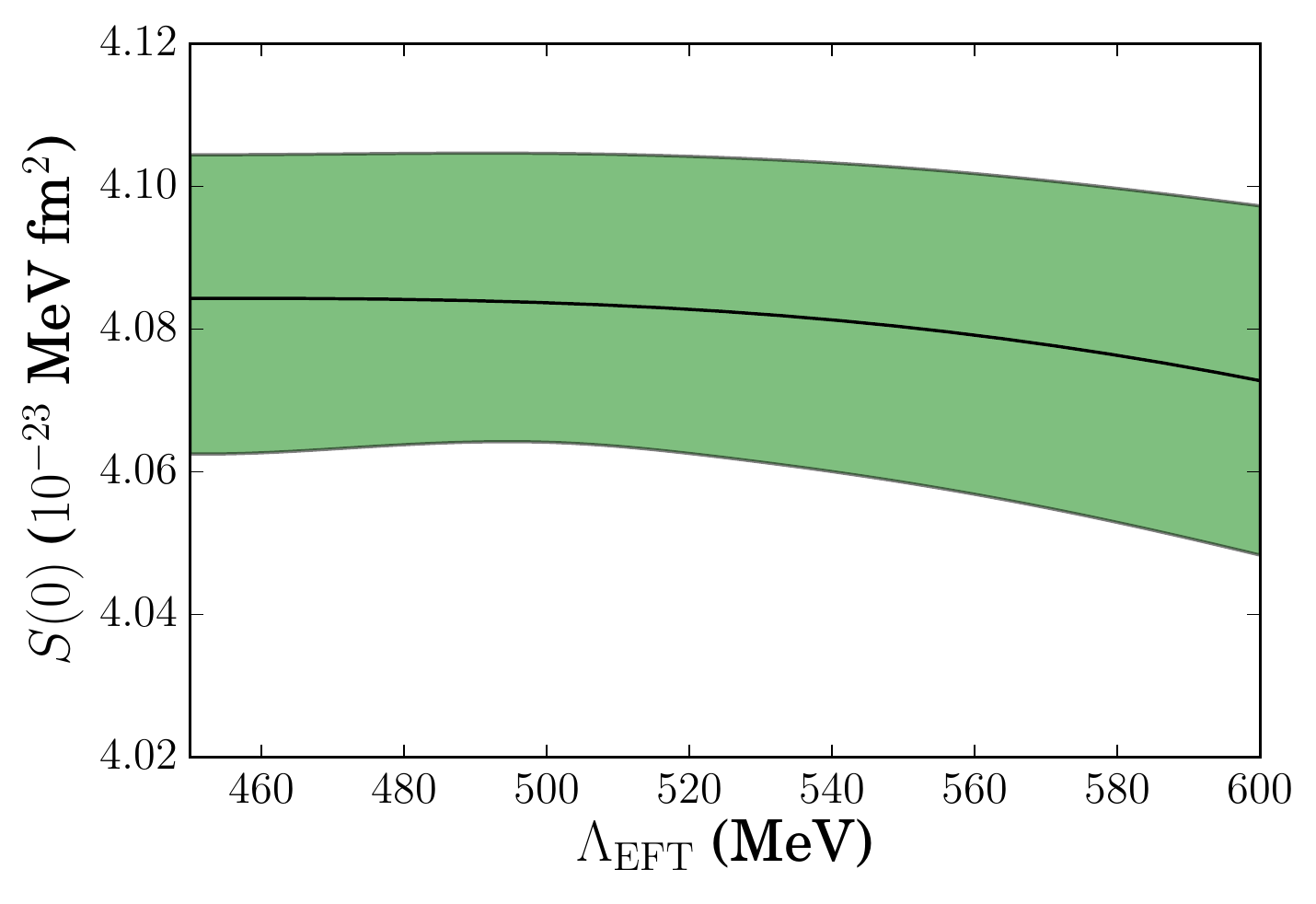}
    \caption{\label{fig:cutoff_variation}\co The green band indicates
      the spread of $S(0)$-values due to variations in $T_{\rm
        Lab}^{\rm max}$ used in the optimization of the NNLO chiral
      force, as well as the propagated statistical uncertainties of
      all LECs and $g_A$, as a function of the cutoff $\Lambda_{\rm
        EFT}$ in the $\chi$EFT. $\Lambda_{\rm EFT}$ was varied between
      450 MeV and 600 MeV in steps of 25 MeV. The cutoff in the
      current and the interaction sectors were always equal to each
      other. This figure demonstrates that the $S$-factor is
      relatively insensitive to reasonable variations in the cutoff.}
\end{figure}

As previously mentioned, systematic uncertainties in the chiral
expansion is probed using the family of 42 different simultaneously
optimized NNLO potentials.  The range of NNLO predictions for $S(0)$,
including the total statistical uncertainty, is shown in
Fig.~\ref{fig:cutoff_variation}. For a given cutoff $\Lambda_{\rm
  EFT}$, the width of the green band indicates the magnitude of all
considered uncertainties. The total error budget is dominated by the
statistical uncertainties in the sub-leading LECs of the chiral
expansion and the axial-vector coupling constant $g_A$. We operate
with currents and interactions that have been optimized simultaneously
and at the same chiral order, which ensures consistent
renormalization.

\subsection{\label{sec:correlationanalysis}Correlation analysis}
\begin{figure}
  \centering
  \includegraphics[width=\columnwidth]{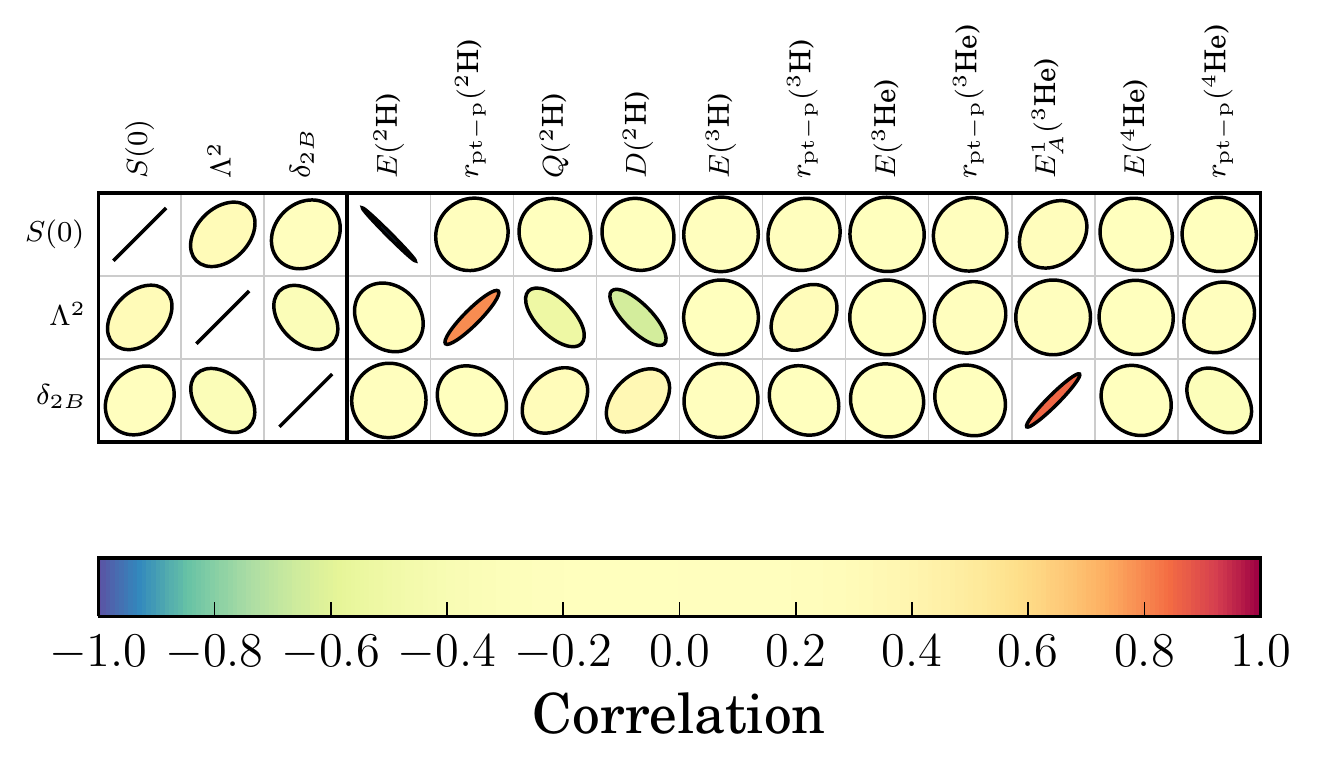} \caption{\label{fig:correlations}\co
    Correlation matrix of the zero-energy $S$-factor ($S(0)$), the squared
    radial wave function overlap ($\Lambda^2$), and the ratio of the
    2B and 1B current matrix elements ($\delta_{\rm 2B}$). We also
    show the correlations between these quantities and the ground
    state energies ($E$), point-proton radii ($r_{\rm pt-p}$) for
    $A=2,3,4$ nuclei as well as the matrix element of the reduced
    axial-vector current ($E_{A}^{1}$) of the triton $\beta$-decay and
    the quadrupole moment ($Q(^2$H)) and $D$-state probability
    ($D(^2$H)) of the deuteron. }
\end{figure}

In addition to the diagonal variances, we also compute the statistical
correlations between all relevant $pp$-fusion quantities and
observables. This study includes masses, radii, and half-lifes of
$A=2,3,4$ nuclei. Correlations can possibly reveal more information,
but this exercise also serves as a sanity check of the entire
uncertainty analysis. We should recover correlations expected from
physical arguments. We employ the Jacobian and covariance matrices of
$A=2,3,4$ observables with respect to the NNLO LECs published in
Ref.~\cite{Carlsson:2015vda} and contract those with the spline
Jacobians extracted in this work. A graphical representation of the
relevant correlations is shown in Fig.~\ref{fig:correlations}. This
particular correlation matrix is based on the NNLO interaction with
$\Lambda_{\rm EFT}=500$ MeV and $T_{\rm Lab}^{\rm max} = 290$ MeV. The
same pattern emerges with any of the 42 different interactions
employed in this work.
As expected from the $Q$-value dependence of the phase space volume, the $S$-factor
strongly anticorrelates with the deuteron ground state energy. It is
noteworthy that the squared radial overlap $\Lambda^2$ of the deuteron
and relative-proton wave functions does not correlate significantly
with $S(0)$. This indicates that the dependence of the $S$-factor on
binding energy indeed occurs predominantly through the phase space. We
also observe that an increase in the deuteron radius would increase the
radial overlap with the proton-proton wave function. The quadrupole
moment of the deuteron and its $D$-state probability anti-correlate with
$\Lambda^2$. Here, it is important to point out that our squared
radial overlap only contains the 1B piece of the current
operator. Thus it only measures the overlap between $S$-wave
components. A smaller $D$-state probability implies a larger
$S$-state probability. Consequently, the anti-correlation between
$\Lambda^2$ and $Q(^2$H)/$D(^2$H) mostly traces the same underlying
$S$-wave component of the deuteron wave function. Finally, we observe
a strong correlation between the strength of the 2B current and the
reduced axial-vector current of the triton $\beta$-decay. In fact, the LEC
$c_D$ plays a dominant role for both currents. In conclusion, we
quantify all expected correlations and confirm that they emerge in our
statistical analysis.

\section{Results and Discussion}
\label{sec:results-discussion}
We have calculated the $pp$-fusion $S$-factor using $\chi$EFT and
carried out a state-of-the-art uncertainty analysis by employing a
family of mathematically optimized chiral potentials at NNLO with
consistently renormalized currents. We focused on the threshold
$S$-factor and have therefore only considered initial $S$-wave $pp$
scattering. To $\mathcal{O}(\alpha)$, we obtain a threshold $S$-factor
\begin{equation}
\label{purecoulomb}
 S(0) = (4.081^{+0.024}_{-0.032}) \times 10^{-23}~{\rm MeV~fm}^2~,
\end{equation}
where we combined, for simplicity, all uncertainties by adding them in
quadrature, and then taking the min/max values of the green band in
Fig~\ref{fig:cutoff_variation}. This error represents all uncertainties
originating from $\chi$EFT, the computational method, and the
statistical extrapolation to obtain the threshold value. The effects
of higher order electromagnetic contributions that are proportional to
$\alpha^2$ remains to be accounted for. These corrections lower the
threshold $S$-factor by about a
percent~\cite{Schiavilla:1998je,Park:1998wq,Marcucci:2013tda}.  From
the energy dependence of these corrections, calculated in
Ref.~\cite{Schiavilla:1998je}, we estimate a 0.84\% reduction in
$S(0)$. The inclusion of these electromagnetic effects leaves the
uncertainties that are due to the strong interaction unchanged, and
the final result becomes
\begin{equation}
\label{eq:finalresult}
S_\mathrm{cor}(0) = (4.047^{+0.024}_{-0.032}) \times 10^{-23}~{\rm MeV~fm}^2~.
\end{equation}

For comparison, the uncertainty presented here is four times larger
than the estimate reported in the pioneering $\chi$EFT calculation in
Ref~\cite{Marcucci:2013tda}.  The comparison of the central values,
however, is not so straightforward since their calculation includes
additional terms in the current operator involving additional LECs,
namely $g_{4S}$ and $g_{4V}$, and the relativistic correction to the
axial one-body current. We estimate the contribution of these missing
terms to be of the order of 0.1\% . 
However, our value given in Eq.~\eqref{eq:finalresult} seems to be
slightly higher than the result obtained by Marcucci {\it et al.} even when these
corrections are considered. This issue could be related to
the infrared convergence of the matrix elements between a bound state
and a scattering state wave function, as mentioned briefly in
Sec.~\ref{sec:wavefunctions}. Additional details will be communicated
through a separate publication~\cite{Acharya-Ekstrom-et-al}.

Our values for the derivatives, $S^{\prime}(0)/S(0)$ and
$S^{\prime\prime}(0)/S(0)$, are, respectively, 10.84(2)~MeV$^{-1}$ and
317.8(13)~MeV$^{-2}$, where the errors also account for variations in
$\Lambda_{\rm EFT}$.

Furthermore, our work displays that great care has to be taken in
order to obtain reliable uncertainty estimates in EFT
calculations. While it had previously been understood that a cutoff
variation is one necessary part in the quantification of
uncertainties, our analysis shows that potentials that use the same
    regulator but are optimized to a different energy range of
    scattering data can lead to different results for electroweak
    matrix elements. This is even more remarkable as the various
    potentials give the same low-energy observables.  We also find
    that the choice of the fit interval is an important consideration
    when using polynomial extrapolation to find the threshold
    $S$-factor. The appropriate interval has to be chosen by comparing
    the statistical errors of the fit for several different fit
    intervals. 
    
In this work, we have only considered 
$S$-wave initial state interaction. In the calculation of Ref.~\cite{Marcucci:2013tda}, 
capture in the $P$-wave channel contributed about 
1\% to $S(E)$ at low $E$, which is roughly the same size as 
the errors stemming from our $\chi$EFT description of the interactions 
and from higher order electromagnetic processes. 
More generally, the effect of higher partial waves on a 
low-energy scattering process has the generic supression 
given by $\delta_l \propto k^{2l+1}$, 
which suggests that the $P$-wave contribution to the fusion cross section formally enters 
at $\mathcal{O}(Q^4)$, which is beyond the order we work to in this paper.

It would be interesting to combine our type of analysis with the
results obtained using pionless EFT. There, the $S$-factor is
expressed as a low-energy expansion that contains only effective range
parameters and a few low-energy constants. Such a parameterization
should be particularly useful in extracting reliable uncertainties for
the energy dependence of the $S$-factor.  Future work also involves
to consider muon-capture by the deuteron. This two-nucleon process
contains information on the $\hat d_R$-term that is strongly tied to the
three-nucleon force. This study could provide additional important
information to constrain the nuclear many-body Hamiltonian.

\begin{acknowledgments}
  We would like to thank Thomas Papenbrock, Doron Gazit, and Laura
  Marcucci for useful discussions.  We are highly obliged to Laura
  Marcucci for sending us data to benchmark our codes with, and for 
  providing valuable comments on the manuscript.  This work
  was supported by the U.S.~Department of
  Energy, Office of Science, Office of Nuclear Physics,  
  under Contract No. DE-AC05-00OR22725, the National Science
  Foundation under Grant No. PHY-1516077, the European Research
  Council under the European Community's Seventh Framework Programme
  (FP7/2007-2013) / ERC Grant No.\ 240603, and the Swedish Foundation
  for International Cooperation in Research and Higher Education
  (STINT, Grant No.\ IG2012-5158).
  Computations were performed on resources provided by the Swedish
  National Infrastructure for Computing (SNIC) at the National
  Supercomputer Centre (NSC).
\end{acknowledgments}

\bibliography{pp}

\begin{thebibliography}{30}
\expandafter\ifx\csname natexlab\endcsname\relax\def\natexlab#1{#1}\fi
\expandafter\ifx\csname bibnamefont\endcsname\relax
  \def\bibnamefont#1{#1}\fi
\expandafter\ifx\csname bibfnamefont\endcsname\relax
  \def\bibfnamefont#1{#1}\fi
\expandafter\ifx\csname citenamefont\endcsname\relax
  \def\citenamefont#1{#1}\fi
\expandafter\ifx\csname url\endcsname\relax
  \def\url#1{\texttt{#1}}\fi
\expandafter\ifx\csname urlprefix\endcsname\relax\def\urlprefix{URL }\fi
\providecommand{\bibinfo}[2]{#2}
\providecommand{\eprint}[2][]{\url{#2}}

\bibitem[{\citenamefont{Adelberger et~al.}(1998)\citenamefont{Adelberger,
  Austin, Bahcall, Balantekin, Bogaert, Brown, Buchmann, Cecil, Champagne,
  de~Braeckeleer et~al.}}]{Adelberger:1998qm}
\bibinfo{author}{\bibfnamefont{E.~G.} \bibnamefont{Adelberger}},
  \bibinfo{author}{\bibfnamefont{S.~M.} \bibnamefont{Austin}},
  \bibinfo{author}{\bibfnamefont{J.~N.} \bibnamefont{Bahcall}},
  \bibinfo{author}{\bibfnamefont{A.~B.} \bibnamefont{Balantekin}},
  \bibinfo{author}{\bibfnamefont{G.}~\bibnamefont{Bogaert}},
  \bibinfo{author}{\bibfnamefont{L.~S.} \bibnamefont{Brown}},
  \bibinfo{author}{\bibfnamefont{L.}~\bibnamefont{Buchmann}},
  \bibinfo{author}{\bibfnamefont{F.~E.} \bibnamefont{Cecil}},
  \bibinfo{author}{\bibfnamefont{A.~E.} \bibnamefont{Champagne}},
  \bibinfo{author}{\bibfnamefont{L.}~\bibnamefont{de~Braeckeleer}},
  \bibnamefont{et~al.}, \bibinfo{journal}{Reviews of Modern Physics}
  \textbf{\bibinfo{volume}{70}}, \bibinfo{pages}{1265} (\bibinfo{year}{1998}).

\bibitem[{\citenamefont{Adelberger et~al.}(2011)\citenamefont{Adelberger,
  Garc\'{\i}a, Robertson, Snover, Balantekin, Heeger, Ramsey-Musolf, Bemmerer,
  Junghans, Bertulani et~al.}}]{Adelberger:2010qa}
\bibinfo{author}{\bibfnamefont{E.~G.} \bibnamefont{Adelberger}},
  \bibinfo{author}{\bibfnamefont{A.}~\bibnamefont{Garc\'{\i}a}},
  \bibinfo{author}{\bibfnamefont{R.~G.~H.} \bibnamefont{Robertson}},
  \bibinfo{author}{\bibfnamefont{K.~A.} \bibnamefont{Snover}},
  \bibinfo{author}{\bibfnamefont{A.~B.} \bibnamefont{Balantekin}},
  \bibinfo{author}{\bibfnamefont{K.}~\bibnamefont{Heeger}},
  \bibinfo{author}{\bibfnamefont{M.~J.} \bibnamefont{Ramsey-Musolf}},
  \bibinfo{author}{\bibfnamefont{D.}~\bibnamefont{Bemmerer}},
  \bibinfo{author}{\bibfnamefont{A.}~\bibnamefont{Junghans}},
  \bibinfo{author}{\bibfnamefont{C.~A.} \bibnamefont{Bertulani}},
  \bibnamefont{et~al.}, \bibinfo{journal}{Reviews of Modern Physics}
  \textbf{\bibinfo{volume}{83}}, \bibinfo{pages}{195} (\bibinfo{year}{2011}).

\bibitem[{\citenamefont{Bethe and Critchfield}(1938)}]{Bethe:1938yy}
\bibinfo{author}{\bibfnamefont{H.~A.} \bibnamefont{Bethe}} \bibnamefont{and}
  \bibinfo{author}{\bibfnamefont{C.~L.} \bibnamefont{Critchfield}},
  \bibinfo{journal}{Physical Review} \textbf{\bibinfo{volume}{54}},
  \bibinfo{pages}{248} (\bibinfo{year}{1938}).

\bibitem[{\citenamefont{{Bahcall} and {May}}(1969)}]{Bahcall:1968wz}
\bibinfo{author}{\bibfnamefont{J.~N.} \bibnamefont{{Bahcall}}}
  \bibnamefont{and} \bibinfo{author}{\bibfnamefont{R.~M.} \bibnamefont{{May}}},
  \bibinfo{journal}{The Astrophysical Journal} \textbf{\bibinfo{volume}{155}}, \bibinfo{pages}{501}
  (\bibinfo{year}{1969}).

\bibitem[{\citenamefont{{Kamionkowski} and
  {Bahcall}}(1994)}]{Kamionkowski:1993fr}
\bibinfo{author}{\bibfnamefont{M.}~\bibnamefont{{Kamionkowski}}}
  \bibnamefont{and} \bibinfo{author}{\bibfnamefont{J.~N.}
  \bibnamefont{{Bahcall}}}, \bibinfo{journal}{The Astrophysical Journal}
  \textbf{\bibinfo{volume}{420}}, \bibinfo{pages}{884} (\bibinfo{year}{1994}).

\bibitem[{\citenamefont{Schiavilla et~al.}(1998)\citenamefont{Schiavilla,
  Stoks, Gl\"ockle, Kamada, Nogga, Carlson, Machleidt, Pandharipande, Wiringa,
  Kievsky et~al.}}]{Schiavilla:1998je}
\bibinfo{author}{\bibfnamefont{R.}~\bibnamefont{Schiavilla}},
  \bibinfo{author}{\bibfnamefont{V.~G.~J.} \bibnamefont{Stoks}},
  \bibinfo{author}{\bibfnamefont{W.}~\bibnamefont{Gl\"ockle}},
  \bibinfo{author}{\bibfnamefont{H.}~\bibnamefont{Kamada}},
  \bibinfo{author}{\bibfnamefont{A.}~\bibnamefont{Nogga}},
  \bibinfo{author}{\bibfnamefont{J.}~\bibnamefont{Carlson}},
  \bibinfo{author}{\bibfnamefont{R.}~\bibnamefont{Machleidt}},
  \bibinfo{author}{\bibfnamefont{V.~R.} \bibnamefont{Pandharipande}},
  \bibinfo{author}{\bibfnamefont{R.~B.} \bibnamefont{Wiringa}},
  \bibinfo{author}{\bibfnamefont{A.}~\bibnamefont{Kievsky}},
  \bibnamefont{et~al.}, \bibinfo{journal}{Physical Review C}
  \textbf{\bibinfo{volume}{58}}, \bibinfo{pages}{1263} (\bibinfo{year}{1998}).

\bibitem[{\citenamefont{Park et~al.}(1998)\citenamefont{Park, Kubodera, Min,
  and Rho}}]{Park:1998wq}
\bibinfo{author}{\bibfnamefont{T.-S.} \bibnamefont{Park}},
  \bibinfo{author}{\bibfnamefont{K.}~\bibnamefont{Kubodera}},
  \bibinfo{author}{\bibfnamefont{D.-P.} \bibnamefont{Min}}, \bibnamefont{and}
  \bibinfo{author}{\bibfnamefont{M.}~\bibnamefont{Rho}}, \bibinfo{journal}{The
  Astrophysical Journal} \textbf{\bibinfo{volume}{507}}, \bibinfo{pages}{443}
  (\bibinfo{year}{1998}).

\bibitem[{\citenamefont{Park et~al.}(2003)\citenamefont{Park, Marcucci,
  Schiavilla, Viviani, Kievsky, Rosati, Kubodera, Min, and Rho}}]{Park:2002yp}
\bibinfo{author}{\bibfnamefont{T.-S.} \bibnamefont{Park}},
  \bibinfo{author}{\bibfnamefont{L.~E.} \bibnamefont{Marcucci}},
  \bibinfo{author}{\bibfnamefont{R.}~\bibnamefont{Schiavilla}},
  \bibinfo{author}{\bibfnamefont{M.}~\bibnamefont{Viviani}},
  \bibinfo{author}{\bibfnamefont{A.}~\bibnamefont{Kievsky}},
  \bibinfo{author}{\bibfnamefont{S.}~\bibnamefont{Rosati}},
  \bibinfo{author}{\bibfnamefont{K.}~\bibnamefont{Kubodera}},
  \bibinfo{author}{\bibfnamefont{D.-P.} \bibnamefont{Min}}, \bibnamefont{and}
  \bibinfo{author}{\bibfnamefont{M.}~\bibnamefont{Rho}},
  \bibinfo{journal}{Physical Review C} \textbf{\bibinfo{volume}{67}},
  \bibinfo{pages}{055206} (\bibinfo{year}{2003}).

\bibitem[{\citenamefont{Marcucci et~al.}(2013)\citenamefont{Marcucci,
  Schiavilla, and Viviani}}]{Marcucci:2013tda}
\bibinfo{author}{\bibfnamefont{L.~E.} \bibnamefont{Marcucci}},
  \bibinfo{author}{\bibfnamefont{R.}~\bibnamefont{Schiavilla}},
  \bibnamefont{and} \bibinfo{author}{\bibfnamefont{M.}~\bibnamefont{Viviani}},
  \bibinfo{journal}{Physical Review Letters} \textbf{\bibinfo{volume}{110}},
  \bibinfo{pages}{192503} (\bibinfo{year}{2013}).

\bibitem[{\citenamefont{Kong and Ravndal}(2001)}]{Kong:2000px}
\bibinfo{author}{\bibfnamefont{X.}~\bibnamefont{Kong}} \bibnamefont{and}
  \bibinfo{author}{\bibfnamefont{F.}~\bibnamefont{Ravndal}},
  \bibinfo{journal}{Physical Review C} \textbf{\bibinfo{volume}{64}},
  \bibinfo{pages}{044002} (\bibinfo{year}{2001}).

\bibitem[{\citenamefont{Butler and Chen}(2001)}]{Butler:2001jj}
\bibinfo{author}{\bibfnamefont{M.}~\bibnamefont{Butler}} \bibnamefont{and}
  \bibinfo{author}{\bibfnamefont{J.-W.} \bibnamefont{Chen}},
  \bibinfo{journal}{Physics Letters B} \textbf{\bibinfo{volume}{520}},
  \bibinfo{pages}{87 } (\bibinfo{year}{2001}).

\bibitem[{\citenamefont{Ando et~al.}(2008)\citenamefont{Ando, Shin, Hyun, Hong,
  and Kubodera}}]{Ando:2008va}
\bibinfo{author}{\bibfnamefont{S.}~\bibnamefont{Ando}},
  \bibinfo{author}{\bibfnamefont{J.}~\bibnamefont{Shin}},
  \bibinfo{author}{\bibfnamefont{C.}~\bibnamefont{Hyun}},
  \bibinfo{author}{\bibfnamefont{S.}~\bibnamefont{Hong}}, \bibnamefont{and}
  \bibinfo{author}{\bibfnamefont{K.}~\bibnamefont{Kubodera}},
  \bibinfo{journal}{Physics Letters B} \textbf{\bibinfo{volume}{668}},
  \bibinfo{pages}{187 } (\bibinfo{year}{2008}).

\bibitem[{\citenamefont{Chen et~al.}(2013)\citenamefont{Chen, Liu, and
  Yu}}]{Chen:2012hm}
\bibinfo{author}{\bibfnamefont{J.-W.} \bibnamefont{Chen}},
  \bibinfo{author}{\bibfnamefont{C.-P.} \bibnamefont{Liu}}, \bibnamefont{and}
  \bibinfo{author}{\bibfnamefont{S.-H.} \bibnamefont{Yu}},
  \bibinfo{journal}{Physics Letters B} \textbf{\bibinfo{volume}{720}},
  \bibinfo{pages}{385 } (\bibinfo{year}{2013}).

\bibitem[{\citenamefont{Ekström et~al.}(2013)}]{Ekstrom:2013kea}
\bibinfo{author}{\bibfnamefont{A.}~\bibnamefont{Ekström}},
\bibinfo{author}{\bibfnamefont{G.} \bibnamefont{Baardsen}},
\bibinfo{author}{\bibfnamefont{C.}~\bibnamefont{Forss{\'e}n}},
\bibinfo{author}{\bibfnamefont{G.} \bibnamefont{Hagen}},
\bibinfo{author}{\bibfnamefont{M.~H.} \bibnamefont{Jensen}},
\bibinfo{author}{\bibfnamefont{G.~R.} \bibnamefont{Jansen}},
\bibinfo{author}{\bibfnamefont{R.}~\bibnamefont{Machleidt}},
\bibinfo{author}{\bibfnamefont{W.} \bibnamefont{Nazarewicz}},
\bibinfo{author}{\bibfnamefont{T.} \bibnamefont{Papenbrock}},
\bibinfo{author}{\bibfnamefont{J.} \bibnamefont{Sarich}},
\bibnamefont{and} \bibinfo{author}{\bibfnamefont{S.~M.} \bibnamefont{Wild}},
 \bibinfo{journal}{Physical Review Letters}
  \textbf{\bibinfo{volume}{110}}, \bibinfo{pages}{192502}
  (\bibinfo{year}{2013}).

\bibitem[{\citenamefont{Navarro~P\'erez
  et~al.}(2014)\citenamefont{Navarro~P\'erez, Amaro, and
  Arriola}}]{navarro2014}
\bibinfo{author}{\bibfnamefont{R.}~\bibnamefont{Navarro~P\'erez}},
  \bibinfo{author}{\bibfnamefont{J.~E.} \bibnamefont{Amaro}}, \bibnamefont{and}
  \bibinfo{author}{\bibfnamefont{E.~R.} \bibnamefont{Arriola}},
  \bibinfo{journal}{Physical Review C} \textbf{\bibinfo{volume}{89}},
  \bibinfo{pages}{024004} (\bibinfo{year}{2014}).

\bibitem[{\citenamefont{Ekstr{\"o}m et~al.}(2015)\citenamefont{Ekstr{\"o}m,
  Carlsson, Wendt, Forss{\'e}n, Jensen, Machleidt, and Wild}}]{Ekstrom2015}
\bibinfo{author}{\bibfnamefont{A.}~\bibnamefont{Ekstr{\"o}m}},
  \bibinfo{author}{\bibfnamefont{B.~D.} \bibnamefont{Carlsson}},
  \bibinfo{author}{\bibfnamefont{K.~A.} \bibnamefont{Wendt}},
  \bibinfo{author}{\bibfnamefont{C.}~\bibnamefont{Forss{\'e}n}},
  \bibinfo{author}{\bibfnamefont{M.~H.} \bibnamefont{Jensen}},
  \bibinfo{author}{\bibfnamefont{R.}~\bibnamefont{Machleidt}},
  \bibnamefont{and} \bibinfo{author}{\bibfnamefont{S.~M.} \bibnamefont{Wild}},
  \bibinfo{journal}{Journal of Physics G} \textbf{\bibinfo{volume}{42}},
  \bibinfo{pages}{034003} (\bibinfo{year}{2015}).

\bibitem[{\citenamefont{P\'erez et~al.}(2015)\citenamefont{P\'erez, Amaro, and
  Arriola}}]{navarro2015b}
\bibinfo{author}{\bibfnamefont{R.} \bibnamefont{Navarro~P\'erez}},
  \bibinfo{author}{\bibfnamefont{J.~E.} \bibnamefont{Amaro}}, \bibnamefont{and}
  \bibinfo{author}{\bibfnamefont{E.~R.} \bibnamefont{Arriola}},
  \bibinfo{journal}{Physical Review C} \textbf{\bibinfo{volume}{91}},
  \bibinfo{pages}{054002} (\bibinfo{year}{2015}).

\bibitem[{\citenamefont{Carlsson et~al.}(2016)\citenamefont{Carlsson,
  Ekstr\"om, Forss\'en, Str\"omberg, Jansen, Lilja, Lindby, Mattsson, and
  Wendt}}]{Carlsson:2015vda}
\bibinfo{author}{\bibfnamefont{B.~D.} \bibnamefont{Carlsson}},
  \bibinfo{author}{\bibfnamefont{A.}~\bibnamefont{Ekstr\"om}},
  \bibinfo{author}{\bibfnamefont{C.}~\bibnamefont{Forss\'en}},
  \bibinfo{author}{\bibfnamefont{D.~F.} \bibnamefont{Str\"omberg}},
  \bibinfo{author}{\bibfnamefont{G.~R.} \bibnamefont{Jansen}},
  \bibinfo{author}{\bibfnamefont{O.}~\bibnamefont{Lilja}},
  \bibinfo{author}{\bibfnamefont{M.}~\bibnamefont{Lindby}},
  \bibinfo{author}{\bibfnamefont{B.~A.} \bibnamefont{Mattsson}},
  \bibnamefont{and} \bibinfo{author}{\bibfnamefont{K.~A.} \bibnamefont{Wendt}},
  \bibinfo{journal}{Physical Review X} \textbf{\bibinfo{volume}{6}},
  \bibinfo{pages}{011019} (\bibinfo{year}{2016}).

\bibitem[{\citenamefont{Feenberg and Trigg}(1950)}]{Feenberg:1950ft}
\bibinfo{author}{\bibfnamefont{E.}~\bibnamefont{Feenberg}} \bibnamefont{and}
  \bibinfo{author}{\bibfnamefont{G.}~\bibnamefont{Trigg}},
  \bibinfo{journal}{Reviews of Modern Physics} \textbf{\bibinfo{volume}{22}},
  \bibinfo{pages}{399} (\bibinfo{year}{1950}).

\bibitem[{\citenamefont{Kurylov et~al.}(2003)\citenamefont{Kurylov,
  Ramsey-Musolf, and Vogel}}]{Kurylov:2002vj}
\bibinfo{author}{\bibfnamefont{A.}~\bibnamefont{Kurylov}},
  \bibinfo{author}{\bibfnamefont{M.~J.} \bibnamefont{Ramsey-Musolf}},
  \bibnamefont{and} \bibinfo{author}{\bibfnamefont{P.}~\bibnamefont{Vogel}},
  \bibinfo{journal}{Physical Review C} \textbf{\bibinfo{volume}{67}},
  \bibinfo{pages}{035502} (\bibinfo{year}{2003}).

\bibitem[{\citenamefont{Walecka}(1995)}]{Walecka:1995mi}
\bibinfo{author}{\bibfnamefont{J.~D.} \bibnamefont{Walecka}}, 
\bibinfo{title}{Theoretical Nuclear and Subnuclear Physics},
  \bibinfo{publisher}{Oxford University Press~(1995)}.

\bibitem[{\citenamefont{Park et~al.}(1996)\citenamefont{Park, Min, and
  Rho}}]{Park:1995pn}
\bibinfo{author}{\bibfnamefont{T.-S.} \bibnamefont{Park}},
  \bibinfo{author}{\bibfnamefont{D.-P.} \bibnamefont{Min}}, \bibnamefont{and}
  \bibinfo{author}{\bibfnamefont{M.}~\bibnamefont{Rho}},
  \bibinfo{journal}{Nuclear Physics A} \textbf{\bibinfo{volume}{596}},
  \bibinfo{pages}{515 } (\bibinfo{year}{1996}).

\bibitem[{\citenamefont{Song et~al.}(2009)\citenamefont{Song, Lazauskas, and
  Park}}]{Song:2008zf}
\bibinfo{author}{\bibfnamefont{Y.-H.} \bibnamefont{Song}},
  \bibinfo{author}{\bibfnamefont{R.}~\bibnamefont{Lazauskas}},
  \bibnamefont{and} \bibinfo{author}{\bibfnamefont{T.-S.} \bibnamefont{Park}},
  \bibinfo{journal}{Physical Review C} \textbf{\bibinfo{volume}{79}},
  \bibinfo{pages}{064002} (\bibinfo{year}{2009}).

\bibitem[{\citenamefont{Men\'endez et~al.}(2011)\citenamefont{Men\'endez,
  Gazit, and Schwenk}}]{Menendez:2011qq}
\bibinfo{author}{\bibfnamefont{J.}~\bibnamefont{Men\'endez}},
  \bibinfo{author}{\bibfnamefont{D.}~\bibnamefont{Gazit}}, \bibnamefont{and}
  \bibinfo{author}{\bibfnamefont{A.}~\bibnamefont{Schwenk}},
  \bibinfo{journal}{Physical Review Letters} \textbf{\bibinfo{volume}{107}},
  \bibinfo{pages}{062501} (\bibinfo{year}{2011}).

\bibitem[{\citenamefont{Tomoda}(1991)}]{Tomoda:1990rs}
\bibinfo{author}{\bibfnamefont{T.}~\bibnamefont{Tomoda}},
  \bibinfo{journal}{Reports on Progress in Physics}
  \textbf{\bibinfo{volume}{54}}, \bibinfo{pages}{53} (\bibinfo{year}{1991}).

\bibitem[{\citenamefont{Vincent and Phatak}(1974)}]{Vincent:1974}
\bibinfo{author}{\bibfnamefont{C.~M.} \bibnamefont{Vincent}} \bibnamefont{and}
  \bibinfo{author}{\bibfnamefont{S.~C.} \bibnamefont{Phatak}},
  \bibinfo{journal}{Physical Review C} \textbf{\bibinfo{volume}{10}},
  \bibinfo{pages}{391} (\bibinfo{year}{1974}).

\bibitem[{\citenamefont{Olive and Group}(2014)}]{1674-1137-38-9-090001}
\bibinfo{author}{\bibfnamefont{K.}~\bibnamefont{Olive}} \bibnamefont{and}
  \bibinfo{author}{\bibfnamefont{P.~D.} \bibnamefont{Group}},
  \bibinfo{journal}{Chinese Physics C} \textbf{\bibinfo{volume}{38}},
  \bibinfo{pages}{090001} (\bibinfo{year}{2014}).

\bibitem[{\citenamefont{Liu et~al.}(2010)\citenamefont{Liu, Mendenhall, Holley,
  Back, Bowles, Broussard, Carr, Clayton, Currie, Filippone
  et~al.}}]{PhysRevLett.105.181803}
\bibinfo{author}{\bibfnamefont{J.}~\bibnamefont{Liu}},
  \bibinfo{author}{\bibfnamefont{M.~P.} \bibnamefont{Mendenhall}},
  \bibinfo{author}{\bibfnamefont{A.~T.} \bibnamefont{Holley}},
  \bibinfo{author}{\bibfnamefont{H.~O.} \bibnamefont{Back}},
  \bibinfo{author}{\bibfnamefont{T.~J.} \bibnamefont{Bowles}},
  \bibinfo{author}{\bibfnamefont{L.~J.} \bibnamefont{Broussard}},
  \bibinfo{author}{\bibfnamefont{R.}~\bibnamefont{Carr}},
  \bibinfo{author}{\bibfnamefont{S.}~\bibnamefont{Clayton}},
  \bibinfo{author}{\bibfnamefont{S.}~\bibnamefont{Currie}},
  \bibinfo{author}{\bibfnamefont{B.~W.} \bibnamefont{Filippone}},
  \bibnamefont{et~al.} (\bibinfo{collaboration}{UCNA Collaboration}),
  \bibinfo{journal}{Physical Review Letters} \textbf{\bibinfo{volume}{105}},
  \bibinfo{pages}{181803} (\bibinfo{year}{2010}).

\bibitem[{\citenamefont{Abele et~al.}(2002)\citenamefont{Abele,
  Astruc~Hoffmann, Bae\ss{}ler, Dubbers, Gl\"uck, M\"uller, Nesvizhevsky,
  Reich, and Zimmer}}]{PhysRevLett.88.211801}
\bibinfo{author}{\bibfnamefont{H.}~\bibnamefont{Abele}},
  \bibinfo{author}{\bibfnamefont{M.}~\bibnamefont{Astruc~Hoffmann}},
  \bibinfo{author}{\bibfnamefont{S.}~\bibnamefont{Bae\ss{}ler}},
  \bibinfo{author}{\bibfnamefont{D.}~\bibnamefont{Dubbers}},
  \bibinfo{author}{\bibfnamefont{F.}~\bibnamefont{Gl\"uck}},
  \bibinfo{author}{\bibfnamefont{U.}~\bibnamefont{M\"uller}},
  \bibinfo{author}{\bibfnamefont{V.}~\bibnamefont{Nesvizhevsky}},
  \bibinfo{author}{\bibfnamefont{J.}~\bibnamefont{Reich}}, \bibnamefont{and}
  \bibinfo{author}{\bibfnamefont{O.}~\bibnamefont{Zimmer}},
  \bibinfo{journal}{Physical Review Letters} \textbf{\bibinfo{volume}{88}},
  \bibinfo{pages}{211801} (\bibinfo{year}{2002}).

\bibitem[{\citenamefont{Acharya}()}]{Acharya-Ekstrom-et-al}
\bibinfo{author}{\bibfnamefont{B.}~\bibnamefont{Acharya}~\bibnamefont{{\it et al.}}}, \bibinfo{journal}{In preparation}.

\end{thebibliography}
\bibliographystyle{apsrev}

\end{document}